\documentclass{aa}  

\usepackage{graphicx}
\usepackage{txfonts}

\usepackage{color}

\newcommand{\reply}[1]{{#1}}
\newcommand{\newtxt}[1]{{#1}}

\begin{document}
\title{Continued activity in P/2013 P5 PANSTARRS 
\thanks{Based on
    observations collected at the European Southern Observatory, La
    Silla, Chile (NTT), program 184.C-1143(H), the Canada France
    Hawai`i Telescope, Mauna Kea, Hawai`i, and the 1.2m telescope on
    Calar Alto, Spain.}
}

\subtitle{The comet that should not be}

 \author{
     O.~R.~Hainaut \inst{1} \and 
     H. Boehnhardt \inst{2} \and
     C. Snodgrass \inst{2} \and
     K.~J.~Meech   \inst{3,4} \and 
     J.~Deller \inst{2,5} \and
     M.~Gillon \inst{6} \and
     E.~Jehin \inst{6} \and
     E.~Kuehrt \inst{7} \and
     S.~C.~Lowry \inst{5} \and
     J.~Manfroid \inst{6} \and
     M.~ Micheli \inst{3} \and
     S.~Mottola \inst{7} \and
     C.~Opitom \inst{6} \and
     J.-B.~Vincent \inst{2} \and
     R.~Wainscoat \inst{3}
}

 \institute{
     European Southern Observatory (ESO), Karl Schwarzschild Stra\ss
     e, 85\,748 Garching bei M\"unchen, Germany -- \email{ohainaut@eso.org}
     \and
     Max Planck Institute for Solar System Research, Justus-von-Liebig-Weg 3, 37077 G\"ottingen, Germany  
     \and
     Institute for Astronomy, University of Hawai`i, 2680 Woodlawn
     Drive, Honolulu, HI 96822, USA 
     \and
     NASA Astrobiology Institute, USA
     \and
     Centre for Astrophysics and Planetary Science, School of Physical
     Sciences, The University of Kent, Canterbury CT2 7NH, UK 
     \and
     Institut d’Astrophysique et de G\'eophysique, Universit\'e de Li\`ege, 
     All\'ee du 6 ao\^ut 17, Sart Tilman, Li\`ege 1, Belgium
     \and
     DLR-German Aerospace Center, Institute of Planetary Research,
     Rutherfordstr. 2, D-12489 Berlin, Germany 
}

   \date{Received 2013 October 17; accepted 2014 January 20}

 \abstract{The object \object{P/2013~P5 PANSTARRS} was discovered in August 2013,
   displaying a cometary tail, but its orbital elements indicated
   that it was a typical member of the inner asteroid Main Belt. We
   monitored the object from 2013 August 30 until 2013 October 05 using the
   CFHT 3.6m telescope (Mauna Kea, HI), the NTT (ESO, La Silla), the
   CA~1.23~m telescope (Calar Alto), the Perkins 1.8m (Lowell) and the
   0.6m TRAPPIST telescope (La Silla).  We measured its nuclear radius to be $r
   \lesssim 0.25-0.29$~km, and its colours $g'-r' = 0.58\pm 0.05$ and
   $r'-i'=0.23\pm 0.06$, typical for an S-class asteroid, as expected
   for an object in the inner asteroid belt and in the vicinity of the
   Flora collisional family. We failed to detect any rotational
   light curve with an amplitude $<0.05$~mag and a double-peaked rotation
   period $<20$~h.  The evolution of the tail during the
   observations was as expected from a dust tail. A detailed
   Finson-Probstein analysis of deep images acquired with the NTT in
   early September and with the CFHT in late September indicated that
   the object was active since at least late January 2013 until the
   time of the latest observations in 2013 September, with at least
   two peaks of activity around 2013 June 14$\pm 10$~d and
   2013 July 22$\pm 3$~d. The changes of activity level and the
   activity peaks were extremely sharp and short, shorter than the
   temporal resolution of our observations ($\sim 1$~d).  The dust
   distribution was similar during these two events, with dust grains
   covering at least the 1--1000$\mu$m range. The total mass ejected
   in grains $<1$~mm was estimated to be $3.0\, 10^6$~kg and $2.6\,
   10^7$~kg around the two activity peaks.
\newtxt{ Rotational disruption cannot be ruled out as the cause of the
  dust ejection.  We also propose that the components of a contact
  binary might gently rub and produce the observed emission.  }
\reply{ Volatile sublimation might also explain what appears as cometary
   activity over a period of 8~months. However, while Main Belt comets
   best explained by ice sublimation are found in the outskirts of the Main
   Belt, where water ice is believed to be able to survive  buried in
   moderately large objects for the  
   age of the solar system deeply,
   the presence of volatiles in an object smaller than 300~m in radius
   would be very surprising in the inner asteroid belt.  }
} 


 \maketitle
%
\section{Introduction}

Since 1996, a dozen  objects have been discover to display dust activity
typical of comets, while they are on orbits typical of Main Belt asteroids (see
\cite{Jew12} for a review), 
\reply{including 
P/2013~P5, which is discussed in this paper}. Their semi-major axes are smaller
than Jupiter's, and their Tisserand parameters are larger than 3,
indicating that they are dynamically decoupled from Jupiter
\citep{Kre80}. For many of them, orbital integrations indicate that they
are long-term residents of the Main Belt (eg. \cite{HJI09} for
238P, \cite{HH+12} for 300163), and simulations even demonstrate that
they very likely formed in situ \citep{Hag09}, which rules out a formation in
the traditional comet reservoirs of the Oort Cloud or the Kuiper belt.

{\bf Long-lasting activity:}
Some of these objects showed dust emission that extended over periods of
months around perihelion ---over 60d for 133P \citep{Boe98}, 100d for
238P \citep{HMP11}, 200d for P/2010~R2, \citep{Mor11}--- suggesting
that volatile ice sublimation (traditional cometary activity) is most
likely the process responsible for lifting the dust from the object
surface.  Two objects were found active at consecutive perihelion
passages: 133P \citep{HJF04} and 238P \citep{HMP11}: this activity
pattern also suggests that ice sublimation is the underlying process that
drives the activity.
Water ice is the only volatile that can survive in Main Belt objects
\citep{PrR09}, and that only as crystalline ice and in minute amounts
\citep{MeS04}.  
\reply{Theoretical considerations  \citep{Sch08}
  indicate that water should be able to survive for the age of the
  solar system even in a fairly small asteroid, provided that it is
  buried under a protective, insulating layer of regolith.
The Main Belt comets (MBCs) from this category are all located in the outskirts of
the Main Belt, where water ice survival is slightly less challenging
\citep{Sch08}.}


\newtxt{Until now, no direct spectroscopic signature from any volatile
  has been detected in MBCs \citep{Sno13}.
These non-detections are not constraining, however: most of them
targeted the CN emission, where the detection limits were 1--2 orders
of magnitude above the expected amount of gas required to lift the
observed dust ---see for instance \citet{Lic11} for observations of
133P and 176P. Furthermore, they relied on the standard CN/H$_2$O ratio
for normal comets. Because Main Belt comets are likely to be completely
depleted of all volatiles but water, this ratio is at best an
optimistic upper limit that makes the non-detection of CN even less
constraining for the presence of water.  }

 

{\bf Short activity:} 
Other active asteroids displayed a very short, impulsive burst of
dust, which slowly dispersed under the solar radiation pressure. The
activity of P/2012 F5 \citep{Ste12, Mor12}, 596 \citep{HYH12, Ish11a,
  Ish11b, Jew11,Mor12}, and P/2010 A2 \citep{Jew10, Sno10, Jew11,
  Hai12, Kim12, Kle13} has been interpreted as the effect of impacts
by small asteroids.

It must be noted that rotational instability is also considered a
possible cause for the dust emission in P/2012 A2 \citep{Aga13} and
possibly P/2012 F5 (Agarwal, private com.). For instance, the YORP
effect \citep{Low07, Tay07, Mar11, JaS11} can accelerate the rotation
of a small object until it reaches the critical period at which the
centrifugal force on the surface is stronger[[[ than the gravity and
tensile strength of the body. Rotational light curves of asteroids
suggest that objects larger than 150~m in diameter are inside the
limit for losing mass at the equator \citep{Pra02}, which suggests that this
process might be in effect.  Because P/2012~A2 is a very small object
\citep[$\sim 100$m radius post-emission, ][ the matter ejected
  amounts to a very small change of the radius]{Jew10, Hai12}, this
is quite plausible. For completeness, the YORP effect is also believed
to be able to cause a catastrophic break-up of the object, a scenario
incompatible with these observations.



\reply{While some still-unknown processes might play a role, }
cometary activity in the outer Main Belt, and impacts and possibly
rotation break-up elsewhere, seem to satisfactorily explain most of
the observed objects. However, there are already some exceptions to
this preliminary scheme: 176P, which was discovered while active in
the outer Belt, did not show any activity on the next perihelion
passage \citep{HH13}, indicating that an extended period of activity
does not guarantee cometary behaviour. However, once activated, each
MBC is likely to remain active only for a certain period of time until
the hypothesised near-surface volatiles are depleted. While 259P was
found to be active in the inner Belt, this object is probably has
arrived recently on its current orbit \citep{Jew09}, which suggests
that it is an interloper.

\citet{Jew12} discussed a list of processes that might
cause activity in asteroids: in addition to those already considered,
electrostatic levitation and radiation-sweeping could in
principle lift small dust grains from small asteroids. However, given
the large number of asteroids and the large amount of dust seen for
many of the MBCs, the proposal that this mechanism is a plausible
cause for activity does not address the question why most of the main belt
asteroids show no dust comae. The other processes, thermal fracture
and thermal dehydration, are not considered to be relevant for P/2013~P5
because its orbit is not close to the Sun.

\subsection{P/2013~P5 (PANSTARRS) }
\begin{figure}

\includegraphics[width=8.8cm]{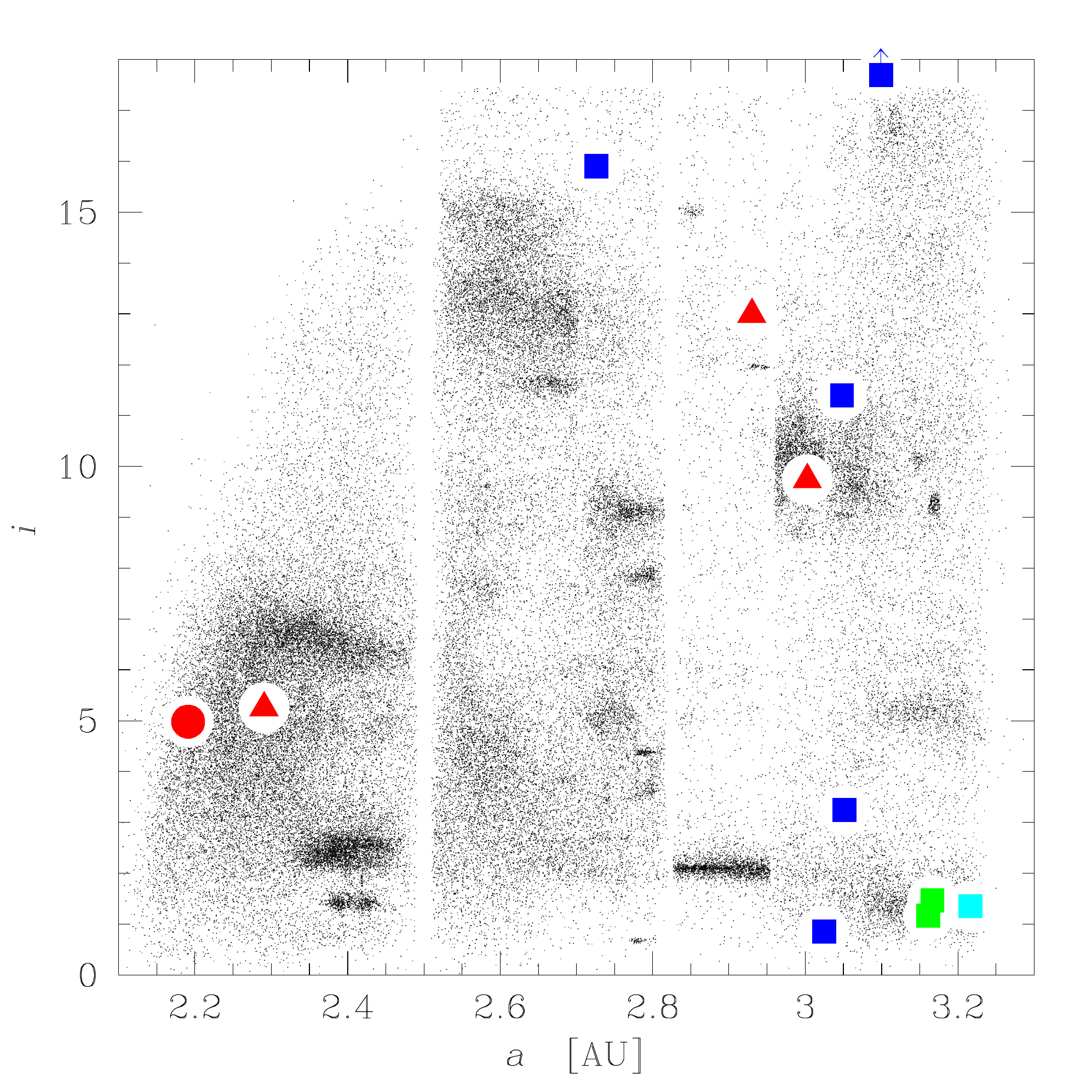}
\caption{\reply{
Orbital elements (inclination vs semi-major axis) of objects in the
asteroid Main Belt.  The small dots represent the background
population of Main Belt asteroids (only numbered ones are represented
for clarity; proper elements were used to put collisional families in
evidence as dense clouds; source: the Asteroid Dynamics Site,
http://hamilton.dm.unipi.it/astdys/).  The known Main Belt comets are
represented as large symbols: squares indicate objects with
extended activity; those that showed activity at more than one
perihelion are plotted in green; 176P, which did not display activity when
returning to perihelion, is marked in cyan.  The inclination of
P/2010~R2 is off scale (i = 21.39); its $a$ is marked by a square with
an arrow at the top of the plot.  Triangles indicate objects that
showed only a very brief activity episode.  P5 is indicated by the
large red circle to the left.
}
}\label{fig:ele}
\end{figure}

The object P/2013~P5 (PANSTARRS) ---hereafter P5--- was discovered on 2013 August 15
with the Pan-STARRS1 telescope on Haleakala, Hawai`i, with a 30$''$
tail \citep{CBET3639}. Follow-up observations showed a tail at least
$90''$ long. Its orbital elements ($a=2.19$~AU, $e=0.117$,
$i=4.98\degr$, marked in Fig.~\ref{fig:ele}) and its Tisserand
parameter $T_J=3.6$ place P5 in the inner asteroid belt, at the edge
of the region occupied by the Flora family. Judging from its position in the
Main Belt, P5 probably is a S-class object. The orbit of P5 has the
smallest semi-major axis of all MBCs and also the lowest orbital
eccentricity.

%


\section{Observations}\label{sec:obs}

\begin{table*}
\caption{Observation log and geometry}     
\label{tab:log}      
\begin{tabular}{llcrrrrrrc}
\hline
Date$^{1}$& Telescope/Instrument& $r^{2}$ & $\Delta^{3}$ &
$\phi^{4}$ & PsAng$^{5}$ &  PsAMV$^{6}$ & Airm.$^{7}$ & Exp.T.$^{8}$ &Filter \\
\hline\hline
2013-Aug-30.367 & CFHT/MegaPrime       &    2.127 & 1.128 &  5.3 & 200.4 & 245.2 & 1.182 &    900  &   r \\ 
2013-Aug-30.380 & CFHT/MegaPrime       &    2.127 & 1.128 &  5.3 & 200.4 & 245.2 & 1.139 &    900  &   g \\ 
2013-Sep-01.153 & NTT/EFOSC2           &    2.125 & 1.124 &  4.7 & 190.9 & 245.1 & 1.247 &   4500  &   R \\ 
2013-Sep-03.250 & NTT/EFOSC2           &    2.122 & 1.120 &  4.2 & 176.9 & 245.0 & 1.264 &  13500  &   R \\ 
2013-Sep-05.342 & CFHT/MegaPrime       &    2.119 & 1.117 &  4.0 & 159.6 & 245.0 & 1.206 &   1800  &   r \\ 
2013-Sep-06.204 & NTT/EFOSC2           &    2.118 & 1.116 &  4.0 & 152.3 & 245.0 & 1.234 &  10200  &   V \\ 
2013-Sep-10.412 & CFHT/MegaPrime       &    2.113 & 1.114 &  5.0 & 122.6 & 244.8 & 1.065 &   1800  &   r \\ 
2013-Sep-11.851 & CA1.23m/DLR-MKIII    &    2.111 & 1.115 &  5.4 & 115.5 & 244.8 & 1.720 &   3300  &   R \\ 
2013-Sep-25.850 & CA1.23m/DLR-MKIII    &    2.092 & 1.145 & 12.2 &  86.3 & 244.4 & 1.665 &   2100  &   R \\ 
2013-Sep-28.264 & CFHT/MegaPrime       &    2.089 & 1.154 & 13.3 &  84.3 & 244.4 & 1.218 &   1048  &   r \\ 
2013-Sep-29.213 & CFHT/MegaPrime       &    2.088 & 1.158 & 13.8 &  83.6 & 244.3 & 1.575 &    540  &   r \\ 
2013-Sep-30.248 & CFHT/MegaPrime       &    2.087 & 1.163 & 13.8 &  82.9 & 244.3 & 1.277 &    540  &   r \\ 
2013-Sep-30.235 & CFHT/MegaPrime       &    2.087 & 1.163 & 13.8 &  82.9 & 244.3 & 1.267 &    540  &   g \\ 
2013-Sep-30.251 & CFHT/MegaPrime       &    2.087 & 1.163 & 13.8 &  82.9 & 244.3 & 1.262 &    540  &   i \\ 
2013-Oct-04.202 & Perkins/PRISM        &    2.082 & 1.182 & 16.0 &  80.7 & 244.2 & 1.245 &   3600  &   R \\
2013-Oct-05.125 & TRAPPIST/PL3041-BB   &    2.080 & 1.188 & 16.5 &  80.2 & 244.2 & 1.160 &  20880 & Rc   \\
\hline
\end{tabular}

{\bf Notes:}
1: UT, mid-exposure; 
2 and 3: helio- and geocentric distances [AU];
4: solar phase angle [degrees];
5 and 6: position angles of the extended Sun--target radius vector,
and the negative of the target's heliocentric velocity vector,
respectively, as seen in the observer's plane of sky [degrees as
  measured N to E]; 
7: airmass; 
8: total exposure time [s].

\end{table*}

Table~\ref{tab:log} lists the observations, and the telescopes and
instruments used are briefly described below. All the observations
were acquired as series of short exposures for which the telescope tracked the
comet. The telescope was moved between each of the exposures by a
small random offset. Twilight
exposures were acquired as flat-fields. Zero-second exposures were
collected during day-time to characterize the bias level. Some
\citet{landolt92} fields were observed, but most of the nights
included here were not photometric. Photometric calibration is
discussed below.

{\bf NTT:} The observations were performed at the ESO
3.56m New Technology Telescope (NTT) on La Silla, with the ESO Faint
Object Spectrograph and Camera (v.2) instrument \citep[EFOSC2, ][]{Buz84,
Sno08}, through Bessel V, and R filters, using the ESO\#40 detector,
a 2k$\times$2k thinned, UV-flooded Loral/Lesser CCD, which was read in a
2$\times$2 bin mode, resulting in 0$\farcs$24 pixels and a 4$\farcm$1 field
of view.  

{\bf CFHT:} The observations were
acquired at the 3.6m Canada France Hawai`i Telescope (CFHT) on Mauna
Kea, with the MegaPrime instrument, using only one of the 36 detectors
in the MegaCam mosaic: the Marconi/EEV CCD42-90 named Auguste, with a
pixel scale of 0$\farcs$187.
The filters used were the Sloan {\it g', r', i'}. 

{\bf CA:} The 1.23~m telescope on Calar Alto was used with the
DLR-KMIII camera equipped with an e2v CCD231-84-NIMO-BI-DD CCD with
4k$\times$4k pixels of 15~$\mu$m. It was used in a 2$\times$2 bin mode,
resulting in a pixel scale of 0$\farcs$628 and a field of view of
$21\farcm 4$. It was used with the Johnson\_R filter.  

{\bf Lowell:} Observations at the Lowell observatory were made with the
Perkins 1.8 m telescope on UT 2013 Oct 4 with the PRISM reimaging
camera (SITE  2k$\times$2k CCD) through the Bessell R band filters under
photometric conditions.  The plate scale of the detector is 0$\farcs$39/pix.

{\bf TRAPPIST:} The comet was observed with the robotic 0.60m TRAPPIST
\citep[TRAnsiting Planets and PlanetesImals Small Telescope;][]{Jeh11}
located at the ESO La Silla Observatory. The camera was a
FLI ProLine PL3041-BB with 2k$\times$2k pixels of 15~$\mu$m.  It was
used with the Cousin R filter and in 2$\times$2 bin mode, resulting in
a pixel scale of 1$\farcs$3 and a field of view of 22$'$.

\section{Data processing}\label{sec:dataproc}

\begin{figure*}
\includegraphics[height=19cm]{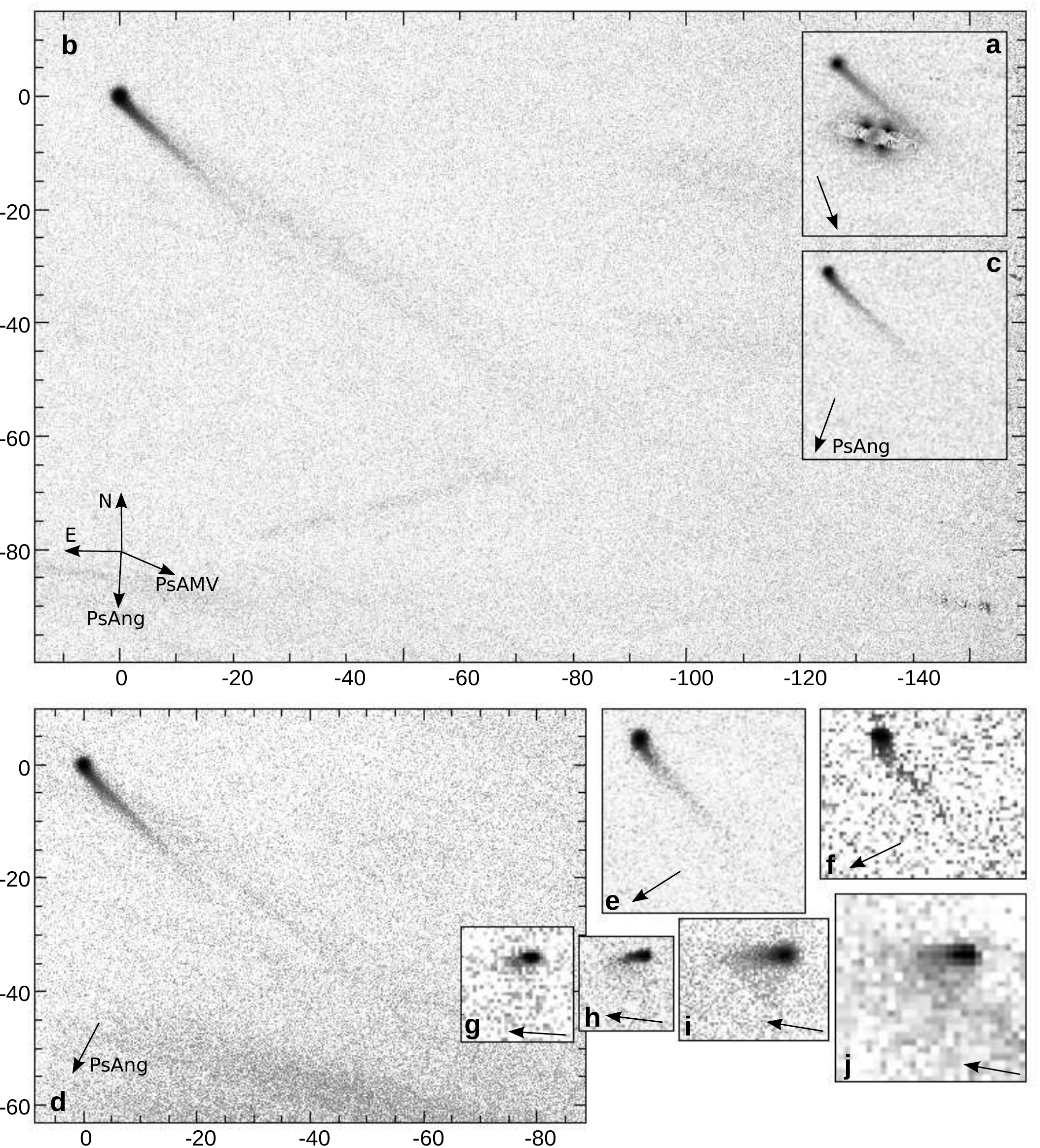}
\caption{P/2013 P5, composite images. \reply{All images have the same
  arcsecond scale (indicated in b and d) and orientation (north
  is up, east is left), to enable direct size and  orientation comparison.
  The composites are star-subtracted, and the grey-scale is negative
  log. The anti-orbital motion (PsAMV) is indicated in b and is
  constant within 1$\degr$; the anti-sun direction (PsAng) is
  indicated in each panel.}
a: CFHT 2013 Aug. 30; 
b: NTT Sep. 03; 
c: CFHT Sep. 05;
d: NTT Sep. 06;
e: CFHT Sep. 10;
f: CA Sep. 11;
g: CA Sep. 25;
h: CFHT Sep. 28+29.
i: Lowell Perkins Oct. 4;
j: TRAPPIST Oct.5.
}
\label{fig:allImg}
\end{figure*}

The data were processed using custom scripts in IRAF and ESO-MIDAS.
The images were corrected for instrumental signature by subtracting a
master bias frame and dividing by a master flat-field frame obtained
from twilight sky images. The offsets between images of a same dataset
were corrected for using series of field stars, and the images stacked
using a median to create a template image of the field, which was
astrometrically calibrated using field stars from the USNO
catalogue. The motion of the comet was compensated for using the
astrometric solution to compute the offsets between expected positions
of the comet from JPL's Horizons ephemerides. The frames were shifted
and then combined using either a median combination or an average
rejecting the pixels deviating from the median value to
remove the stars and background objects as well as detector defects
and cosmic-ray hits. For datasets with enough frames for the template
not to show residuals from the comet, the field template image was
subtracted from the individual frames to remove the background
objects. When needed, the resulting images for each dataset were
flipped to set them to the standard orientation. These final images
are presented in Fig.~\ref{fig:allImg}.

\begin{figure*}
\includegraphics[width=17.5cm]{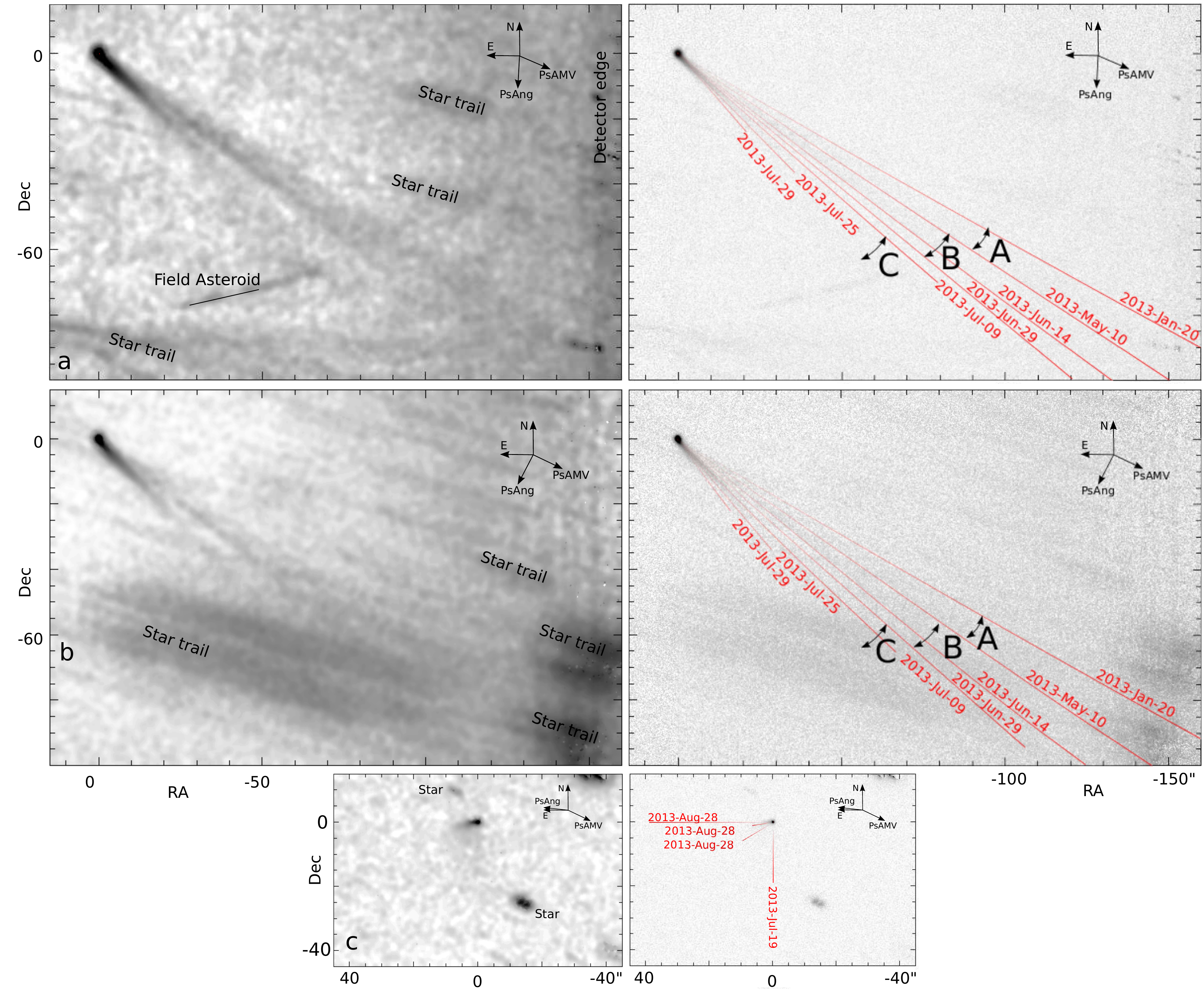}
\caption{\reply{P/2013 P5, composite images enhanced for low
    surface-brightness features (left, see text) and original
    composites (right, from Fig.~\ref{fig:allImg}). The direction of
  the anti-sun (PsAng) and of the anti orbital motion (PsAMV) are
  indicated, axes are labelled in arcsec. The lines and labels mark the
  streamers listed in the text; the dates indicate the epoch at which
  the dust along these lines was released. These images are
    available electronically at the CDS}.
a: 13\,500s $R$, NTT, 2013-Sep-03. 
b: 10\,200s $V$, NTT, 2013-Sep-06. 
c: 1\,588s $r'$, CFHT, 2013-Sep-28+29.
}
\label{fig:fp}
\end{figure*}

\section{Analysis}\label{sub:analysis}
\subsection{Object characterization}
\subsubsection{Light curve}
\begin{figure}
a.\includegraphics[width=8.8cm]{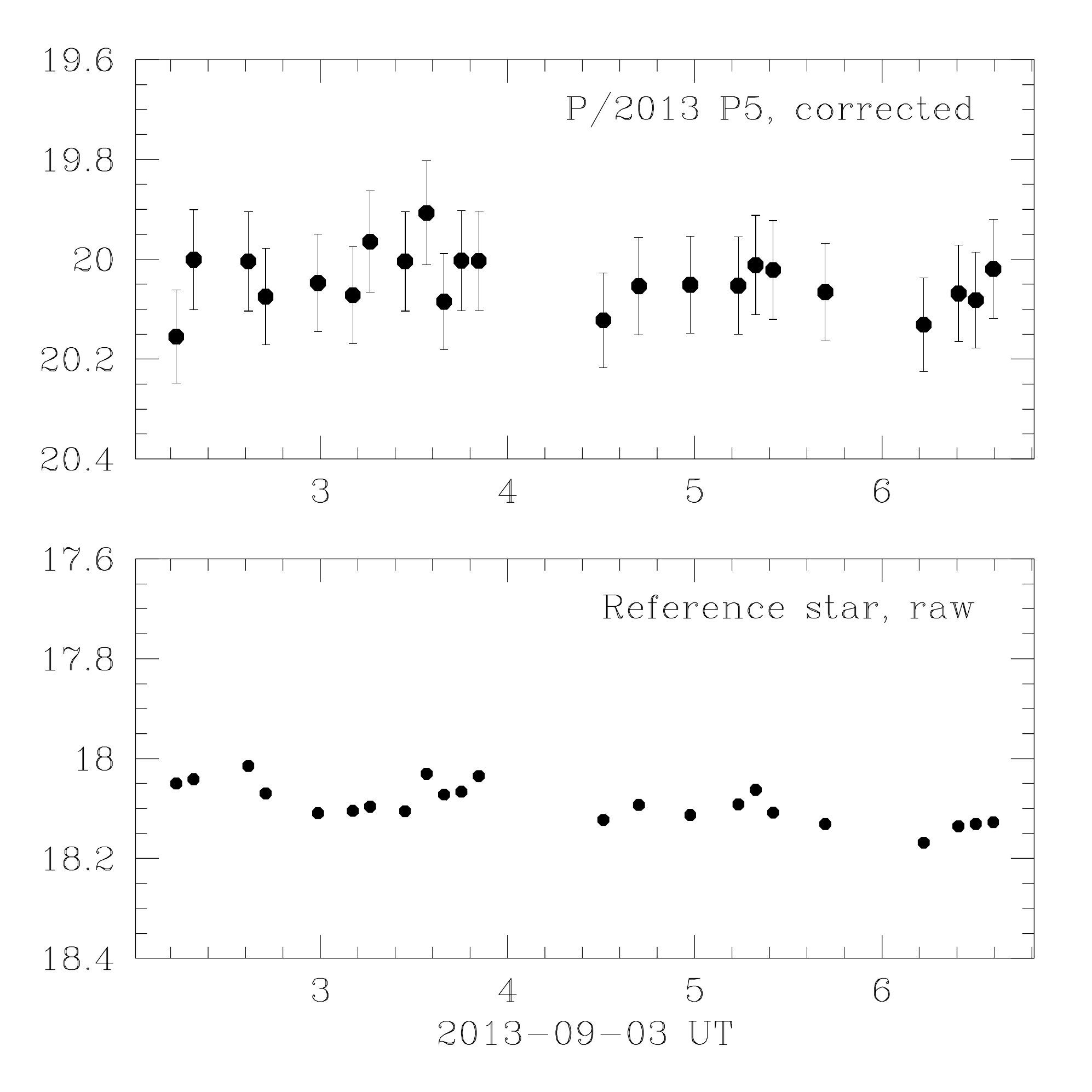}
b.\includegraphics[width=8.8cm]{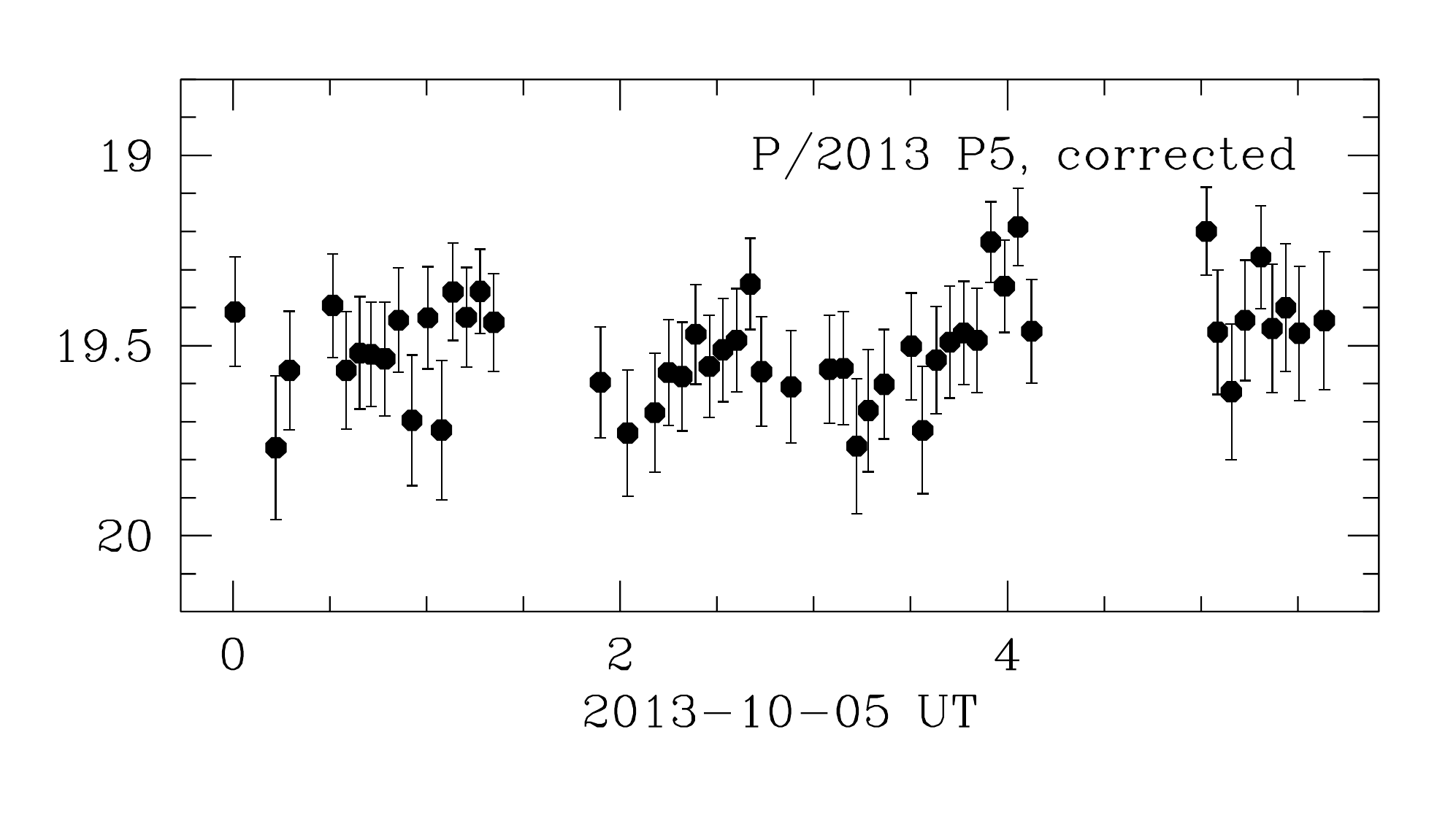}
\caption{a. Light curve of the object (top panel) and of a reference star
   (bottom) observed with the NTT. The object magnitudes were corrected using
  the variations  
   of the reference stars. The night was not perfectly photometric, so
   a zero   point from another night was used.  The reference star
   magnitudes are not corrected for extinction and are shown to
   illustrate that the extinction did not vary by much.  
b. Light curve of the object observed with TRAPPIST.
}
\label{fig:lc}
\end{figure}

The data from the NTT on 2013 September 03 cover more than 4h~40m and were
acquired with a fairly clear sky, even if not photometric; those from
TRAPPIST on 2013 October 05 were acquired under photometric conditions and
cover 6h.  They constitute the best dataset to detect a rotational
light curve. The flux of the object and of a series of non-saturated
stars visible in most of the frames was measured in a series of
concentric apertures centred on the object; the background was
estimated from a much wider annulus, rejecting from the average the
values of outlying pixels. Comparing the light curves of these
reference stars, we decided to pick the 5$''$ aperture for the NTT,
and 4$\farcs$5 (3 pixels) for TRAPPIST.
%
On the NTT, the reference stars show variations of 0.15~mag caused by
the variable extinction, while the TRAPPIST data are stable at the
percent-level.  The magnitude of the object was measured relative to
reference stars.  For the NTT, a photometric zero point $ZP_R=25.86$
was used, obtained a few nights before the observations. The zero
point for the TRAPPIST images was obtained using 79 PPMXL stars in the
field, and tied to star 1417242133114009170 ($\alpha$ 22:20:25.256,
$\delta$ -1:17:14.64). Fig. \ref{fig:lc} displays the light curves.
Gaps are present in the light curves at the time intervals when the
object was too close to a background object.  The light curves were
searched for periodic signal in the 0.1 to 3h range using the 
phase dispersion minimization
method implemented in Peranso (CBABelgium.com). The periodograms do
not show any significant signal; P5 does not show any sign of any
rotational light curve\reply{: both datasets are compatible with
  a sine-wave shaped light curve with a half-amplitude $< 0.05$~mag (NTT) and
  $< 0.14$~mag (TRAPPIST), for light curve periods shorter than twice
  the time span of the data series. Because most asteroid light curve are
  double peaked, this corresponds to a rotational period $<18.5$~h
  (NTT) and $<20$~h (TRAPPIST).}

The reason for this lack of rotational signature can be that the amplitude is too
small to be detected, either because the cross-section of the nucleus
does not change by more than a few percent (either because the nucleus
is not very elongated, or because we are seeing it close to pole-on),
because of the (fairly constant) contamination by the dust dilutes the
nucleus light curve, because the rotation period is longer than
the duration of the observations, or because of a combination of these
factors.

\subsubsection{Colours}
On 2013 September 30, the CFHT acquired three frames in each of the $g'$,
$r'$, and $i'$ filters under photometric conditions. The magnitude of
the object was measured in the co-added images in each filter using a
series of concentric apertures ranging from 1 to 10'' in diameter; the
residual sky background was estimated as the median value of an aperture $\sim
30''$ wide, manually selected on each frame in a region near
the object and free of background sources.  To minimize the
dust contamination by the coma, the instrumental magnitudes measured
in the $3''$ diameter aperture were kept and corrected for the
difference between that aperture and the total flux of an artificial
star with the same FWHM as the field stars. The instrumental
magnitudes were then converted into standard $ugriz$ magnitudes using
the conversion listed in the MegaCam manual, including the standard
photometric solutions. For reference, the BVRI colours were also
computed using the relations in \citet{Fuk96}. They are listed in
Table~\ref{tab:mag}.

\begin{table}
\caption{P5 magnitudes in the AB system and colours from CFHT,
  2013-Sep-30}      
\label{tab:mag}
\begin{tabular}{lll}
\hline
$g'=21.23\pm0.05$ & $g'-r'=0.58\pm0.05$  \\
$r'=20.66\pm0.05$ & $r'-i'=0.23\pm0.06$  \\
$i'=20.43\pm0.06$ &                      \\
 & $B-V=0.77\pm0.03$\\
 & $V-R=0.47\pm0.03$\\
 & $R-I=0.47\pm0.03$ \\

\hline
\end{tabular}
\end{table}

\begin{table}
\caption{Colours of P5 compared with those of other classes of objects.}      
\label{tab:col}
\begin{tabular}{lll}
\hline
Object or class & $g'-r'$          &  $r'-i'$ \\
\hline
\hline
P5              & 0.58$\pm$0.05    & 0.23$\pm$0.06 \\
\hline
Solar$^1$       & 0.400            & 0.109 \\
C-class$^1$     & 0.432            & 0.114 \\
D-class$^1$     & 0.527            & 0.214 \\
S-class$^1$     & 0.600            & 0.196 \\
Short-period comets$^2$ & 0.658    & 0.222 \\
\hline
\end{tabular}

1 \citet{Fit11}; 2 \citet{HBP12}
\end{table}

\reply{In the colour table of the taxonomic classes in the {\it
    ugriz} AB magnitude system from \cite{Fit11}, the colours of P5 are
  compatible with those of the S-class.}  D-class, which is slightly
less red than P5 (see Table~\ref{tab:col}), is a slightly poorer
match. It is also worth noting that 133P and 176P \citep{Lic11}, and
238P \citep{HJI09}, \reply{i.e. the Main Belt comets whose activity patterns
are compatible with sublimation-driven activity}, have colours typical of
C-class asteroids ($g'-r'=0.432$, $r'-i'=0.114$), also significantly
less red than those of P5. For completeness, the average $g'-r'$ of 
short-period comets is redder than those of P5 \citep{HBP12}. The position of P5 in
the inner Main Belt supports the hypothesis that it is am member of the S-class.

\subsubsection{Nucleus size}

The CFHT AB magnitudes were converted into the Vega system using the
relations in the MegaCam manual. Together with the NTT magnitude, they
were transformed in absolute magnitude using the geometric information
from Table~\ref{tab:log}. The solar magnitudes used were $r'=-26.95$
\citep{Ive01} and $R=-27.07$ \citep{Pec13}. Using the
average albedo for S-class asteroids, $p=0.23$ \citep{Dem13}, but
without correcting for the (unknown) solar phase correction, the
absolute magnitudes were converted into nuclear radii and are listed in
Table~\ref{tab:rad}. Because the measurements are contaminated by the
resolved coma, these are upper limits.

\begin{table}
\caption{Magnitudes, absolute magnitude (not corrected for solar
  phase effects) and radius of the P5 nucleus.}
\label{tab:rad}
\begin{tabular}{llll}
\hline
Dataset        & Magnitude          & $M(1,1,\alpha)$ & Radius [km] \\
\hline
\hline
2013-Sep-03 & $R=20.0 \pm 0.1$ & $18.1 \pm 0.1$ & $<0.29 \pm 0.02$ \\
2013-Sep-29& $r'_{\rm Vega}=20.49\pm0.05$ & $18.56 \pm 0.05$ &
$0.25 \pm 0.01$\\
\hline
\end{tabular}
\end{table}

\subsubsection{Tail}

\paragraph{Measurements}~\\
\newtxt{{\bf Geometry and nature of the tail:} Figure \ref{fig:allImg}
  shows the evolution of the appearance of P5 during a month around
  opposition, while the viewing geometry changed dramatically, with
  the anti-solar direction spanning over 120$\degr$. Concurrently with
  that change of geometry, the main tail orientation and opening angle
  evolved rapidly, overall as would be expected from a dust tail under
  the effect of radiation pressure.}

\reply{Furthermore, a hypothetical plasma tail would have had very
  different appearances in the deep NTT R and V images (Fig.~2b and d,
  because the main H$_2$O$^+$ band (700--770~nm) is well within the
  transmission bandwidth of the R filter while it is outside the V
  pass-band), and the orientation of a plasma tail would project
  roughly along the north-south direction (that is, perpendicular to
  the solar wind velocity and opposite to the direction of the
  interplanetary electric field, see Hansen et al. 2007). We can
  therefore firmly rule out that the tail of P5 is plasma, and below,
  we consider only a dust tail.  }

\reply{{\bf Identification and measurements of the tail features:} We
  performed a detailed analysis of our deepest images. The first step
  was to identify and measure recognizable morphological features,
  such as possible jets, fans, and streamers.  }

\reply{{\bf Method:}
The images were explored visually, adjusting the display look-up table
and stretch function to reveal low contrast features. Furthermore,
different methods where applied to numerically enhance the very faint
surface brightness features: smoothing and spatial median filtering,
as well as adaptive smoothing.  For the standard smoothing and median
filtering, the filter size used ranged from a few to a few tens of
pixels. The adaptive filtering used is based on the filter/adaptive
command in ESO-MIDAS command, which works as follows: the local
signal-to-noise ratio (S/N) is evaluated at different scale-lengths
via the Laplace H-transform. The smallest order at which the local S/N
reaches a value of 3 sets the local resolution of the filter, and the
signal in that filter is kept. In short, this method smooths the image
with the variable filter size, ensuring that the S/N of each resulting
pixel is 3 or more. Results from the adaptive filtering are displayed
in Fig.~\ref{fig:fp}. 
The composites and their enhanced
versions were also converted into polar coordinates, with the pole on
the nucleus of P5. In these images, radial features in the comet appear as
parallel bands.}

\reply{{\bf Reality of the features:} 
We are confident that all the features listed below are real: 
\begin{itemize}
\item When a feature was detected in one enhanced version of an image, it was
  checked for in others, and in the original composite. Only the
  features that could be confirmed were retained.
\item While the surface brightness of some feature is barely above the
  local background at the pixel level, the total S/N of the streamers
  is fairly high. For instance, the flux from the {\it
    faintest} streamer of the 2013 September 06 image (streamer B, described
  below) was integrated in the original composite over a length of
  40$''$ and over its width, leading to a S/N of 17. The significance
  of the whole streamer is even higher.
\item The features measured on the deep 2013 September 03 composite are recognized
  also in the deep images of 2013 September 06, in spite of the different geometry.
\item As additional a posteriori argument: the Finson-Probstein
  analysis described below leads to the same dust emission date for a
  given feature measured in different images. This also further
  confirms the validity of the hypothesis of the dust nature of the
  tail.
\item A recent article by \citet{Jew13}, which became available while
  the present paper was being reviewed, reported observations of
  similar streamers in P5 images acquired with HST. While these images
  are sensitive only to the very inner part of the object, the
  morphology and position of the streamers are similar to those
  reported here. This is discussed below in more detail.
\end{itemize}
}

\reply{{\bf Precision of the measurements:} 
The position angle (PA) of the features was measured with respect to
the nucleus. Repeating the measurement at various positions on the
feature, the precision of the PA is estimated at $\pm 1\degr$. The
length  of the features was also measured; again, repeating
the measurements, the precision on the length is estimated at $\pm
1''$ to several arcseconds depending on the feature.
}

~\\

\reply{\bf Description of the features:}

{\bf NTT, 2013 Sep. 3 and 6:}
P5 displays a wide dust tail, fanning about 20$\degr$ in the south-west
quadrant, about 40--60$\degr$ from the anti-solar direction. Its length
can be traced over 2$\farcm$5, and appears shorter at decreasing
PA. Three main structures are identified; their lengths ($L$) and PAs
are listed in Table \ref{tab:tailNTT} and are marked in Fig.~\ref{fig:fp}:
\begin{itemize}
\item[A]  Weak, uniform streamer, with a surface brightness slowly
decreasing over PA from 240 to 235$\degr$. The borderlines are labelled
A1 and A2.
\item[B] Narrow streamer, brighter than A. It is limited by B1 (=A2)
  and B2; the brighness peak is Bmax at PA$\sim 223\degr$.
\item[C] Fairly wide, bright streamer, whose surface brightness
  steeply decreases at PA$\sim 230-220\degr$. Its maximum is labelled
  Cmax, and its borderlines are C1 (far from the nucleus) and B2
  (close to the nucleus), and C2. 
\end{itemize}

{\bf CFHT, 2013 Sep. 28 and 29:}
The tail is in the general antisolar direction, brightest at PA$\sim
99\degr$, and not detected beyond PA$<90\degr$; the enhanced image shows
a very faint and diffuse tail toward PA$\sim 180\degr$. No clear
streamer was identified. The visible extension of the tail was
measured at various PAs, which are reported in Table
\ref{tab:tailCFH}.

We only measured the orientation of the tail for the other epochs.

\begin{table*}
\caption{Streamers in the tail of P5: measurements and FP analysis
  results for 2013 Sep. 3 and 6. The uncertainty of the position angle
  measurements is about    1 $\degr$. The epochs are relative to 
  perihelion (2014 April 15).} 
\label{tab:tailNTT}
\begin{tabular}{p{3cm}p{3cm}p{3cm}p{3cm}p{3cm}}
\hline
Tail features & \multicolumn{2}{c}{2013-Sep-3.25} & \multicolumn{2}{c}{2013-Sep-6.20}\\
    & \multicolumn{2}{c}{$t_{\rm obs} = -224.53$~d}  & \multicolumn{2}{c}{$t_{\rm obs} = -221.58$~d}\\
\hline Borderline and/or Maxima & Measurement & FP analysis & Measurements & FP analysis \\  
\hline
\hline
A$_1$ & PA$=240.5\degr$  \newline $L \sim 163''$ 
        & PA$=240\degr$ \newline $\beta_{\rm max}=0.0033$ \newline $t_{\rm em}=-450\pm 20$~d \newline 2013-Jan-20
            & PA$=240.5\degr$  \newline$L>33''$ 
               & PA$=240.5\degr$  \newline$\beta_{\rm max} >0.0001$ \newline $t_{\rm em}= -450\pm 20$~d  \newline2013-Jan-20\\
\hline
A$_2$=B$_1$ & PA$=236\degr$  \newline$L>108''$ 
         & PA$=236\degr$  \newline$\beta_{\rm max}>0.0033$  \newline$t_{\rm em}=-340\pm 10$~d  \newline2013-May-10
            & PA$=235\degr$  \newline$L>33''$ 
               & PA$=236\degr$  \newline$\beta_{\rm  max}>0.0010$ \newline $t_{\rm em}=-340$~d  \newline 2013-May-10 \\
\hline
$B_{\rm max}$ & PA$=233.5\degr$  \newline$L \sim 108''$
               & PA$=333.5\degr$  \newline$\beta_{\rm max}= 0.009$  \newline $t_{\rm em}=-305\pm 10$~d  \newline 2013-Jun-14
                  & PA$=233 \degr$ \newline $L \sim 99''$ 
                    & PA$=332.5\degr$  \newline$\beta_{\rm max}=0.0083$  \newline$t_{\rm em}=-305\pm 10$~d \newline 2013-Jun-14 \\
\hline
B$_2$ & PA$=231\degr$ \newline $L>50''$ 
           & PA$=231\degr$ \newline $\beta_{\rm max} >0.005 $  \newline$t_{\rm em}= -290 \pm 10$~d \newline 2013-Jun-29 
              & PA$=230\degr$  \newline$L>42''$ 
                 & PA$=230\degr$  \newline$\beta_{\rm max} >0.0058$  \newline$t_{\rm em}=-290 \pm 10$~d \newline 2013-Jun-29 \\
\hline
C$_1$  & PA$=229\degr$ \newline $L>100''$ 
           & PA$= 229\degr$ \newline $\beta_{\rm max} >0.008$ \newline $t_{\rm em}=-280 \pm 5$~d  \newline2013-Jul-09 
               & -- 
                   & \\
\hline
C$_{\rm max}$ & PA$=225\degr$ \newline $L=50''$ 
               & PA=$225\degr$ \newline $\beta_{\rm max}=0.005$  \newline $t_{\rm em}=-267.5 \pm 3$~d  \newline 2013-Jul-22
                 & PA$=223\degr$ \newline $L> 31''$ 
                    & PA$=223\degr$  \newline $\beta_{\rm max}= .019$  \newline $t_{\rm em}=-267.5 \pm 3$~d  \newline 2013-Jul-22 \\
\hline
C$_2$   &  PA$=222\degr$ \newline $L >25''$ 
            & PA$=222\degr$  \newline $\beta_{\rm max} >0.023 $  \newline $t_{\rm em}=-260 \pm 3$~d  \newline 2013-Jul-29 
               & PA$=217\degr$ \newline  $L>31''$ 
                  & PA$=218\degr$  \newline $\beta_{\rm max} >0.023$  \newline $t_{\rm em}=-260 \pm 3$~d \newline 2013-Jul-29 \\
\hline
\end{tabular}
\end{table*}
\begin{table}
\caption{Streamers in the tail of P5: measurements and FP analysis
  results for 2013 Sep. 28--29, $t_{\rm obs}=-198.53$~d. The uncertainty of the position angle
  measurements is about  1 $\degr$. The epoch are relative to
  perihelion (2014 April 15).} 
\label{tab:tailCFH}
\begin{tabular}{p{2.5cm}p{2cm}p{3cm}}
\hline
Borderline and/or Maxima & Measurement & FP analysis \\
           & 
                  & 
                        \\
\hline
\hline
North borderline 
  &  PA$=90\degr$    
    & Youngest dust  \newline $t_{\rm em}>-200$~d \newline 2013-Sep-27 \\
Brightest line 
  & PA$=99\degr$ \newline  $L>6''$  
     &$PA=99\degr$ \newline  $\beta_{\rm max} > 0.01$\newline  $t_{\rm em}<-228$d \newline 2013-Aug-29 \\
Longest east-ward extension 
  & PA=$119\degr$ \newline $L \sim 13''$ 
    & PA=$120\degr$ \newline $\beta_{\rm max}=0.0125$   \newline $t_{\rm em}=-238 \pm 3$~d \newline 2013-Aug-20 \\
Longest south-ward extension 
   & PA$=180\degr$ \newline $L \sim 20''$ 
     & PA$=180\degr$\newline  $\beta_{\rm max}= 0.008$ \newline $t_{\rm em}=-269\pm2$~d\newline 2013-Jul-20 \\
\hline
\end{tabular}
\end{table}

\reply{\paragraph{Analysis\\}}

\reply{\bf The Finson-Probstein method:} 
\reply{We used the \citet{FP68} method (hereafter FP) to analyse the
  position and extent of the tail and of the streamers}.  The FP
method assumes that dust grains are released from the nucleus with
zero velocity at an epoch $t_{\rm em}$, and computes their position at
the time of observations $t_{\rm obs}$, taking into account the solar
radiation pressure and gravity. The ratio of these two forces is
$\beta$, which is related to the size and density of the dust grain by
\begin{equation}
\beta = 5.740 \times 10^{-4} \frac{Q_{\rm pr}}{\rho a},
\end{equation}
where $a$ is the grain radius [m] and $\rho$ its density
[kg~m$^{-3}$]; $Q_{\rm pr}$ is the radiation pressure efficiency,
whose typical value is 1--2 depending on the material scattering
properties \reply{\citep{FP68}}. This relation is valid only for particles
larger than the observation wavelength.  In the analysis below, we worked
with $\rho = 3000$~kg~m$^{-3}$ ---the average density of 11 Parthenope
and 20 Massalia, two S-type asteroids \citep{Bri+02}. Of course, the
grains could be porous and therefore have a lower density; the mass
estimates below would then need to be scaled down accordingly.

A line connecting grains of different
$\beta$ emitted at a given $t_{\rm em}$ is a {\em synchrone}, and a line
connecting all grains of a given $\beta$, emitted at different
$t_{\rm em}$, is a {\em syndyne}.  \reply{The FP method does not
  have free parameters: a combination of emission time and force ratio
  will lead to a unique position in the image plane. Whether a
  position in the image plane corresponds to a unique combination of 
 $t_{\rm em}$ and  $\beta$  or a multiple one (or no solution) depends
  on the viewing geometry alone. In the case of P5, we were lucky to
  have a very favourable geometry that nicely spread the syndynes and
  synchrones.}

\reply{\bf Results:}
For the shallow images, the orientation of the tail is found to evolve
quickly from PA$\sim 240\degr$ to $90\degr$ as the object passes
through opposition, as expected from the FP models computed for these
epochs. 

For the deep images, the resulting synchrones appear essentially as
straight lines for emission times about a month before
observation. The dust grains are sorted linearly ---in a very good
approximation--- along the synchrones according to their $\beta$
ratio. From scanning the values of $t_{\rm em}$ and $\beta$, we obtained the
value that matched the PA and length of the streamers and the
corresponding uncertainty ranges; the best fits are reported in
Tables~\ref{tab:tailNTT} and \ref{tab:tailCFH}, \newtxt{and the
  corresponding synchrones are drawn in Fig.~\ref{fig:fp}.
It is worth noting that the emission epochs for the various streamers
measured in the two independent NTT images perfectly match in spite of
the different geometry. Because these epochs are coming straight out
of the FP code, without any adjustable parameter, this additionally
confirms our hypothesis that the tail is composed of dust grains.
}

Because the FP method considers dust grains emitted with zero velocity, the
resulting tail model is a perfectly flat structure in the orbital
plane of the object.  The structures in the object's observed tail are
blurred by two phenomena: the atmospheric seeing, which degrades the
resolution of the image, and the actual dust grain emission velocity
and direction distribution. More observations of the comet when the
Earth crosses the orbital plane would give a direct constraint on the
out-of-plane velocity of the grains.  In the meantime, we can estimate
a limit from the dispersion of streamer C$_{\rm max}$, the sharpest
feature in the tail of P5, emitted 40~d before the observations. If we
consider that its entire width ($\sim 5''$) is caused by dust emission
velocity, this leads to a velocity in the plane of sky of
1.2~m~s$^{-1}$. For comparison, the escape velocity (using the radius
and density discussed above) is $v_e \sim 0.4$~m~$^{-1}$, and a
\citet{Bob54} ejection velocity, assuming the dust is dragged by
sublimating water vapour, is $v_B \sim 500$~m~s$^{-1}$.

It is noted that in the deep NTT exposure of both epochs of early 2013
Sep., there is a tiny cluster of knots with enhanced brightness seen
in streamer B. These knots are at 34, 42 and 53$''$ distance from the
nucleus on 2013 September 03.25 and on 34, 43, and 54$''$ on
2013 September 06.2. These distances are equivalent to $\beta$ = 0.0029,
0.0036 and 0.0045, respectively. Since the area in the images where
these knots are found also contains remnant signals from the removal
of background objects, it remains open whether or not the knots
represent spikes in the distribution of the $\beta$ ratio. At least,
the results found for both observing epochs are consistent with each
other, allowing the interpretation of spikes in the size distribution
of the dust grains. Note that larger chunks ---or even sub-nuclei---
ejected with $\sim 0$ velocity would be located very near to the main
nucleus. Their presence in this region of the tail would imply a
significant ejection velocity, while the whole tail structure is
compatible with 0~velocity.

\begin{figure}
\includegraphics[width=8.8cm]{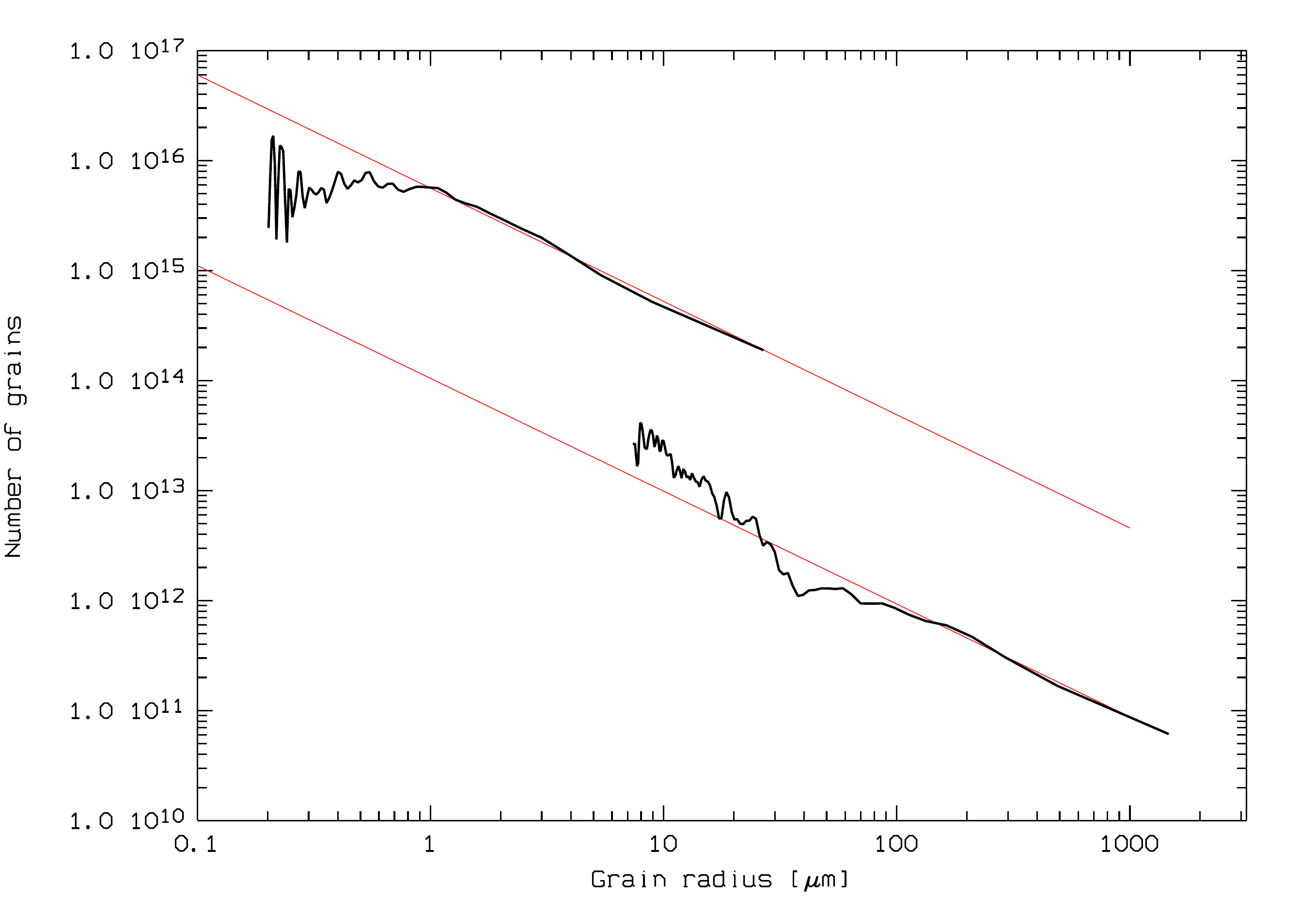}
\caption{\reply{Dust grain distribution (thick line) measured along
    streamers B (bottom) and C (top). The conversion from $\beta$ to
    grain size assumed $\rho = 3000$~kg~m$^{-2}$. In both cases, the
    left end of the distribution is affected by the low
    signal-to-noise ratio in the outskirt of the tail; the right end
    of the distribution was truncated when reaching the seeing disk of
    the nucleus.  The narrow lines are linear fits to the
    distributions where they are not affected by the noise.}}
\label{fig:dust}
\end{figure}

\reply{\bf Quantification of the mass-loss:}
Following the method described in detail in \citet{Hai12}, we
estimated the quantity of dust in the streamers, focusing on the
regions around the two brightest lines, $B_{\rm max}$ and $C_{\rm
  max}$. The distance to the nucleus along one of these lines is
converted into a value for $\beta$ using the output of the FP program,
and into a particle radius $a$ assuming a grain density
$\rho=3000$~kg~m$^{-3}$ and a radiation pressure efficiency $Q_{\rm
  pr}=1$ in Eq.~1. The deep $R$ image from 2013 September 03 was rotated to
align the line that was measured along the $x$-axis of the image, and the
flux measured in a series of $2\times 4''$ or $\times 8''$ apertures
covering the line ($4''$ covers most of the bright region along $B_{\rm
  max}$ and $C_{\rm max}$, $8''$ was used to include the full B
streamer). The flux in each box was converted into a number of
particles, using $p=0.23$, $R_{\sun}=-27.07$, $ZP_R=-25.86$, and the
geometric parameters from Table~\ref{tab:log}. Figure~\ref{fig:dust}
shows the resulting distributions along $B_{\rm max}$ and $C_{\rm
  max}$, together with a linear fit to the part of the distribution
with the best signal-to-noise ratio. The left parts of the curves are
dominated by the noise in the faintest part of the tail, and the right
part is truncated at 2” from the nucleus, because its PSF dominates the
flux. The smaller dust grains of B have left the field of view, while
the large grains of C are lost in the glare of the PSF of the nucleus. The
deviation of the left part of the curve from a line can be fully
accounted for by background objects and variations in the sky
background. Moreover, this is in the range of sizes as small as the
wavelength of observations, so Eq.~1 will not be reliable
anymore. While the absolute number of particles in each streamer is
different, their slopes are very similar, suggesting that the same 
grain distribution was emitted in the B and C streamers.

Integrating the mass of the dust grains in $B_{\rm max}$ leads to a
total mass $m \sim 5.3\, 10^5$ kg, for grains smaller than
1.5~mm. Integrating over the similar distribution measured on the full
$B$ streamer (with a wider aperture) leads to $m \sim 3.0\,
10^6$~kg. As the integrated area corresponds to about 50~days of
activity, this leads to an average mass loss rate of 0.7~kg~s$^{-1}$.

Similarly, integrating the visible grains over $C_{\rm max}$ give $m
\sim 1.0\, 10^4$~kg. As the slopes of the two distributions are very
similar over their different ranges, it is reasonable to consider the
C streamer extends similarly to grains of 1.5~mm; the mass for grain
smaller than 1.5~mm would then scale to $2.6\, 10^7$~kg.  The $C_{\rm
  max}$ peak of activity lasted less than 20~d, so the mass loss
rate was at least 15~kg~s$^{-1}$.




\reply{\bf Interpretation -- activity scenario:}
In the NTT images from early 2013 Sep., the FP analysis indicates that
P5 released dust since late Jan. 2013 (A1) until at least a few days before the
observations (dust extending beyond C2), with a peak on 2013 June 14
(B$_{\rm max}$), and a second peak on
2013 July 22 (C$_{\rm max}$). 
The CFHT images from late Sep. also confirm the dust
production extends until that epoch. Because of the completely
changed geometry during the CFHT observations, and because the images
do not reach very low surface-brightness levels, the streamers
detected on the NTT images are diluted over a larger area and too
faint to be detected. The brightest C streamer appears as the longest
extension of the tail in the CFHT data. This also explains why the
$\beta_{\rm max}$ ratios measured on the CFHT images are lower than
those from the NTT data.

\reply{Based on the FP analysis of the tail and the streamers in the tail, we
conclude that
\begin{itemize}
\item The dust emission started ---or increased from an undetectable
  level--- around 2013 January 20 (marked by A1), 450~d before perihelion,
  at $r=2.35$~AU; based on the angular width of the A1 borderline,
  this increase happened within less than 20~d. This period of dust
  emission lasted for about 110~d (until A2). The fairly constant
  surface-brightness of the streamer between A1 and A2 suggests a
  fairly stable dust release rate. Only grains larger than $\sim
  100\mu$m are visible (for a density $\rho=3000$~kg~m$^{-2}$);
  smaller grains have been pushed beyond the field of view.
\item Around 2013 May 10$\pm 10$ (B1), the activity started to
  increase toward a maximum that took place on 2013 June 14$\pm 10$
  ($r=2.23$AU, marked by B$_{\rm max}$), and then dropped until
  2013 July 09$\pm 5$ (B2 borderline); during this phase, the object
  ejected grains with $\beta$ up to 0.008 (about 10$\mu$m).
\item During the period from 2013 June 29 (B2) until 9 July (C1), the
  surface brightness was much lower, suggesting a lower dust release rate.
\item The activity then increased rapidly, reaching a peak on
  2013 July 22 ($r=2.18$AU, C$_{\rm max}$), i.e. a few weeks before the
  object was discovered. This last streamer is fully within the FoV,
  with $\beta$ up to 0.023, ie micron-sized grains. The C streamer is
  only marginally resolved in width, suggesting this emission episode
  was short. We suggest that this brutal increase contributed to the
  discovery of the object.
\item Both the NTT and CFHT data indicate that the dust release
  continued well into the respective observing epoch, i.e. for CFHT
  until late 2013 Sep.
\end{itemize}
}
\newtxt{
All the changes of activity level are sharp, unresolved in our data,
suggesting that they occurred during short periods of time.
}

In conclusion, P5 was continuously releasing dust during at least the
8 months after late Jan. 2013 showing ---until now--- three to four short
episodes of higher dust emission. Its activity still continued at the
time of the last observations, on 2013 September 29. It has released dust
grains in the 1--1000~$\mu$m range (assuming a density
$\rho=3000$~kg~m$^{-3}$), at low velocities ($<1.2$~m~s$^{-1}$).

\reply{During the refereeing process of this paper, \citet{Jew13}
  published a report of P5 observations acquired with the Hubble Space
  Telescope (HST) on 2013 September 10 and 23. The spatial resolution of the
  HST images is much finer than that of our images, which made the Hubble
  data much more sensitive to unresolved sources ---such as very
  narrow streamers. However, the collecting power of the NTT and
  CFHT, combined with the very long total exposure time of our
  observations allowed us to reach fainter surface brightnesses for
  extended objects, for which high resolution is of no benefit. In
  summary, the two data-sets are complementary. The HST images reveal
  a set of narrow streamers that Jewitt et al. analysed with a method
  similar to that exposed here; they also concluded that each streamer
  might be associated with a single date of dust ejection. The first
  streamer they detected corresponds to mid-April 2013, i.e. HST
  failed to detect the earlier late-January emission of our streamer
  A. They also reported a set of two streamers with emission dates on
  2013 July 18 and 24, which we observe, blended, as the peak of
  streamer~B. Overall, our observations are fully compatible, our deep
  NTT images extending the beginning of the dust emission to an
  earlier time than
  HST's, and  Hubble's high resolution shows that the streamers
  and their edges are even sharper than visible on our
  images.}

\section{Discussion}\label{sub:disc}

\reply{Which  process is responsible for lifting the dust from the
  nucleus of P5?}

An impact, by definition extremely short-lived, cannot reproduce the
observed dust release extending over months, with its increasing dust
production and intermediate maxima.
\newtxt{
Moreover, the dust morphology is very different from that of
impact-triggered emissions, such as that in P/2010~A2 \citep{Hai12} or
in (596) Scheila \citep{Ish11a}. We consider a series of impacts
hitting P5 several times during 2013 as implausible. 
}
A collision could have exposed a hypothetical underlying ice layer
whose sublimation would continue the dust-lifting after the
impact. However, a dust production maximum would be expected at the
beginning of the sequence, which is not observed. More importantly,
the presence of volatile ice would be needed in this scenario; this is
discussed below.

This object has the lowest semi-major axis of all {Main Belt}
active 
asteroids. Its radius is smaller than 300~m. Thermal models developed for
the study of P/2010 A2 \citep[on a similar orbit, but with a radius
  of about 100~m,][]{Hai12} showed that an S-class object would be fully
depleted of water ice within a few $\times 10^7$
years. \reply{Furthermore, P5 might be a member of the Flora
  collisional family and an S-type asteroid; as noted by \citet{Jew13},
  these objects reflect metamorphism to temperatures in the $\sim$ 800
  to 960$\degr$C range. The presence of water ice is therefore unlikely.}

The eccentricity of P5 is the lowest of all the Main Belt
comets, so that the incident solar radiation at perihelion is only 60\%
higher than at aphelion. While the dependency of the sublimation rate is
very steep with temperature \citep{MeS04}, Fig.~5 from
\cite{Jew12} suggests that the change in dust production rate is only of
the order of 2.

\reply{Nevertheless, volatile sublimation (if present) might explain
  the observed dust release: the activity started as the comet moved
  closer to the sun, with some irregularity and short outbursts, but
  with an overall increase. The size and distribution of the lifted
  particles is also as expected for cometary activity.
}

Electrostatic processes and radiation-sweeping would not be able to
lift the large dust grains observed: \citet{Jew12} indicated a $a_{\rm
  max}$ of about 1$\mu$m for an object with the characteristics of P5 (see
his Fig. 5), i.e. much smaller the size of the observed dust grains.

Rotational disruption is a possible explanation, too: while the
rotation of P5
was not detected,  a fast rotating object cannot be ruled
out. Furthermore, the object is small, i.e. in the range that can be
efficiently spun up by the YORP effect ---although the efficiency of the
YORP effect is also strongly tied to the shape of the object, which is
unknown. It is possible that it
reached its critical rotation frequency, causing shedding of particles
from its equator and in a plane, which would mark the beginning of the
activity period.  Observations from different geometries (before and
after orbital-plane crossing) may be able to constrain the 3D geometry of the
tails. Since the YORP effect also drives small asteroids towards
preferred pole directions, this can be considered. Together with a
pole reconstruction from future nucleus light curves, this could
prove or dispute this theory. Of course, these light curves should also reveal
the rotation rate of the nucleus directly, which will also prove or
dispute it.
\newtxt{
For a spherical object, the critical spin can be estimated by setting the
equatorial velocity at the escape velocity $v_e = \sqrt{ 2GM/r }$,
leading to $P_{\rm Crit}  \sim 1.3$--2.3~h for a
density in the 1000--3000~kg~m$^{-3}$ range.
}
The initial ejection of material can be followed by a landslide-like
re-organization of the rubble pile, leading to further dust release
\citep[see][for numerical simulations]{Ric05}. While there is, to our
knowledge, no information on the duration of such a rubble-pile
re-organization, it seems reasonable that it would last much longer
than an impact, but 8 months seems a long time. The initial YORP-induced
disruption could have induced movement leading to exposed sub-surface
ices whose sublimation could have taken over the extended activity
---this last scenario would again imply the presence of volatile ice in the
object.

\newtxt{ Alternatively, the observed dust emission pattern might be
  caused by the two components of a contact or near-contact binary
  object gently rubbing each other. In this scenario, we are
  witnessing the formation of a bilobate asteroid from the low-speed
  merger of a binary pair, as described in recent analytical solutions
  by \citet{Tay13}: before contact is firmly established between the
  components, they are expected to wobble slightly around their
  elongation axis. As the orbit decays, there will be a time when the
  extremities of the two objects come into contact, at very low
  velocity. This would cause some surface break-up and liberation of
  dust. Once the contact is established, it is also likely that the
  two components will settle over an extended period of time, also
  releasing dust. Considering that the total photometric cross-section
  of the object is caused by two spherical objects, they would have
  radii $\sim 250/\sqrt{2} = 180$~m. Setting the object on contact
  orbit, Kepler's third law indicates the rotation/revolution period
  of the system would be in the range of 6.7--11.6~h (for a density in
  the range of 1000-3000~kg~m$^{-3}$); prolate objects will have a slower
  revolution. The total volume of dust ejected (around $10^4$~m$^3$
  from the above-discussed mass estimate) would correspond to the
  shaving of a $\sim$3~m cap from the 180~m radius objects.
  Furthermore, because the system is very elongated, the light curve would
  be expected to have a very large amplitude.  As discussed in Section
  4.1.1., the light curves presented in here do not show any variation,
  but this can be explained by the strong contamination by the dust.
  Again, future nucleus light curves may be able to prove or dispute this theory.\\ }

In summary, our observations of P/2013~P5, from late August until late
2013 September, indicate that the object has emitted dust at least since
late 2013 January, i.e. 450~d before perihelion. While the activity seems
continuous and overall increasing, there were at least two episodes of
more intense activity, around 2013 June 14 and July 22. The dust size
distribution appears to be similar during these events, with dust grains
at least in the range from 1~$\mu$m to 1~mm (larger grains may be
present, but cannot be measured in our data).  The dust mass ejected
during these two peaks of activity amounted to $\sim 3.0\, 10^6$~kg
and $\sim 2.6\, 10^7$~kg, which corresponds to average mass-loss rates
of 0.7 and 15~kg~s$^{-1}$.

\newtxt{Rotational disruption and a rubbing contact binary are plausible
  processes to explain the observed dust release; future observations
  can either prove or disprove this. While the presence of volatile ice in P5
  would be challenging, theoretical studies and models suggest that
  some water could have survived even in the inner asteroid belt. With
  the discovery of the Main Belt comets, models to explain activity
  for objects occupying the outer belt initially suggested that ice
  could remain buried deep, and that having a high obliquity would support
   volatile survival. }

 As more of these enigmatic objects are discovered, it seems more
 likely that not only were there many volatiles in many areas of the
 asteroid belt in the early solar system, but that they may still be
 there.  The dynamical landscape of the young solar system is changing
 as new models try to explain the architecture of our solar system
 \citep{Tsi05, Wal11} and as disk models and observations try to
 refine the location of the water-ice line during the epoch of planet
 building.  Additionally, new dust-releasing processes are being
 uncovered and explored. It is clear that only a detailed study of
 these interesting objects coupled with dynamical models and other
 constraints will help us understand the conditions in the early solar
 system.

\begin{acknowledgements}

Based on observations obtained with MegaPrime MegaCam, a joint project
of CFHT and CEA/DAPNIA, at the Canada-France-Hawaii Telescope (CFHT)
which is operated by the National Research Council (NRC) of Canada,
the Institut National des Science de l'Univers of the Centre National
de la Recherche Scientifique (CNRS) of France, and the University of
Hawaii. 

Based on observations made with ESO Telescopes at the La Silla Paranal
Observatory under programme ID 184.C-1143(H). 

The German-Spanish Astronomical Center at Calar Alto is operated
jointly by the Max-Planck-Institut für Astronomie (MPIA) in
Heidelberg, Germany, and the Instituto de Astrofísica de Andalucía
(CSIC) in Granada/Spain.

TRAPPIST is a project funded by the Belgian Fund for Scientific
Research (Fonds de la Recherche Scientifique, F.R.S FNRS) under grant
FRFC 2.5.594.09.F, with the participation of the Swiss National
Science Fundation (SNF). E. Jehin and M. Gillon are FNRS Research
Associates, J. Manfroid is Research Director FNRS.  C. Opitom thanks
the Belgian FNRS for funding her PhD thesis.

This material is based upon work supported by the National Aeronautics
and Space Administration through the NASA Astrobiology Institute under
Cooperative Agreement No. NNA09DA77A issued through the Office of
Space Science, and is in addition supported by grant \#~1010059 from
the National Science Foundation.  

CS received funding from the European Union Seventh Framework
Programme (FP7/2007-2013) under grant agreement no. 268421.

Image processing in this paper has been performed, in part, using IRAF
\cite{IRAF}. This software is distributed by the National Optical
Astronomy Observatories, which is operated by the Association of
Universities for Research in Astronomy, Inc., under cooperative
agreement with the National Science Foundation, USA.

Image processing in this paper has been performed, in part, using
ESO-MIDAS (version 10SEPpl1). This software was developed and is
distributed by the European Southern Observatory.

We wish the anonymous referee a prompt recovery from his long and
debilitating illness.

\end{acknowledgements}

\bibliographystyle{apalike} 
\bibliography{mnemonic,P2013P5}

\begin{thebibliography}{}

\bibitem[{Agarwal} et~al., 2013]{Aga13}
{Agarwal}, J., {Jewitt}, D., and {Weaver}, H. (2013).
\newblock {Dynamics of Large Fragments in the Tail of Active Asteroid P/2010
  A2}.
\newblock {\em \apj}, 769:46.

\bibitem[{Bobrovnikoff}, 1954]{Bob54}
{Bobrovnikoff}, N.~T. (1954).
\newblock {Reports of observations 1953-1954: Perkins Observatory-Physical
  properties of comets}.
\newblock {\em \aj}, 59:356--358.

\bibitem[{Boehnhardt} et~al., 1998]{Boe98}
{Boehnhardt}, H., {Sekanina}, Z., {Fiedler}, A., {Rauer}, H., {Schulz}, R., and
  {Tozzi}, G. (1998).
\newblock {Impact-Induced Activity of the Asteroid-Comet P/1996N2 Elst-Pizarro:
  Yes or No?}
\newblock {\em Highlights of Astronomy}, 11:233.

\bibitem[{Bolin} et~al., 2013]{CBET3639}
{Bolin}, B., {Denneau}, L., {Micheli}, L., {Winscoat}, W., {Tholen}, D.~J., ,
  {Lister}, T., and {Williams}, G.~V. (2013).
\newblock {Comet P/2013 P5 (Panstarrs)}.
\newblock {\em Central Bureau Electronic Telegrams}, 3639:1.

\bibitem[{Britt} et~al., 2002]{Bri+02}
{Britt}, D.~T., {Yeomans}, D., {Housen}, K., and {Consolmagno}, G. (2002).
\newblock {Asteroid Density, Porosity, and Structure}.
\newblock {\em Asteroids III}, pages 485--500.

\bibitem[{Buzzoni} et~al., 1984]{Buz84}
{Buzzoni}, B., {Delabre}, B., {Dekker}, H., {Dodorico}, S., {Enard}, D.,
  {Focardi}, P., {Gustafsson}, B., {Nees}, W., {Paureau}, J., and {Reiss}, R.
  (1984).
\newblock {The ESO Faint Object Spectrograph and Camera (EFOSC)}.
\newblock {\em The Messenger}, 38:9--13.

\bibitem[{DeMeo} and {Carry}, 2013]{Dem13}
{DeMeo}, F.~E. and {Carry}, B. (2013).
\newblock {The taxonomic distribution of asteroids from multi-filter all-sky
  photometric surveys}.
\newblock {\em \icarus}, 226:723--741.

\bibitem[{Finson} and {Probstein}, 1968]{FP68}
{Finson}, M. and {Probstein}, R. (1968).
\newblock {A theory of dust comets. 1. Model and equations}.
\newblock {\em \apj}, 154:327--380.

\bibitem[{Fitzsimmons}, 2011]{Fit11}
{Fitzsimmons}, A. (2011).
\newblock Asteroid colours in the ps1 photometric system.
\newblock {\em PS1 technical report, Pan-STARRS}, pages
  http://ps1sc.ifa.hawaii.edu/ PS1wiki/ images/
  Fitzsimmons.TheoreticalAsteroidColours--v3.pdf.

\bibitem[{Fukugita} et~al., 1996]{Fuk96}
{Fukugita}, M., {Ichikawa}, T., {Gunn}, J.~E., {Doi}, M., {Shimasaku}, K., and
  {Schneider}, D.~P. (1996).
\newblock {The Sloan Digital Sky Survey Photometric System}.
\newblock {\em \aj}, 111:1748--+.

\bibitem[{Haghighipour}, 2009]{Hag09}
{Haghighipour}, N. (2009).
\newblock {Dynamics, Origin, and Activation of Main Belt Comets}.
\newblock In {\em ``Icy Bodies of the Solar System,'' Proc. IAU Symp. 263, J.
  Fernandez et al., Ed.}

\bibitem[{Hainaut} et~al., 2012a]{HBP12}
{Hainaut}, O.~R., {Boehnhardt}, H., and {Protopapa}, S. (2012a).
\newblock {Colours of minor bodies in the outer solar system. II. A statistical
  analysis revisited}.
\newblock {\em \aap}, 546:A115.

\bibitem[{Hainaut} et~al., 2012b]{Hai12}
{Hainaut}, O.~R., {Kleyna}, J., {Sarid}, G., {Hermalyn}, B., {Zenn}, A.,
  {Meech}, K.~J., {Schultz}, P.~H., {Hsieh}, H., {Trancho}, G.,
  {Pittichov{\'a}}, J., and {Yang}, B. (2012b).
\newblock {P/2010 A2 LINEAR. I. An impact in the asteroid main belt}.
\newblock {\em \aap}, 537:A69.

\bibitem[{Hsieh}, 2013]{HH13}
{Hsieh}, H.~H. (2013).
\newblock {Search for the Return of Activity in Active Asteroid 176P/LINEAR}.
\newblock In {\em AAS/Division for Planetary Sciences Meeting Abstracts},
  volume~45 of {\em AAS/Division for Planetary Sciences Meeting Abstracts},
  page 413.30.

\bibitem[{Hsieh} et~al., 2009]{HJI09}
{Hsieh}, H.~H., {Jewitt}, D., and {Fern{\'a}ndez}, Y.~R. (2009).
\newblock {Albedos of Main-Belt Comets 133P/Elst-Pizarro and 176P/LINEAR}.
\newblock {\em \apjl}, 694:L111--L114.

\bibitem[{Hsieh} et~al., 2004]{HJF04}
{Hsieh}, H.~H., {Jewitt}, D.~C., and {Fern{\'a}ndez}, Y.~R. (2004).
\newblock {The Strange Case of 133P/Elst-Pizarro: A Comet among the Asteroids}.
\newblock {\em \aj}, 127:2997--3017.

\bibitem[{Hsieh} et~al., 2011]{HMP11}
{Hsieh}, H.~H., {Meech}, K.~J., and {Pittichov{\'a}}, J. (2011).
\newblock {Main-belt Comet 238P/Read Revisited}.
\newblock {\em \apjl}, 736:L18.

\bibitem[{Hsieh} et~al., 2012a]{HYH12}
{Hsieh}, H.~H., {Yang}, B., and {Haghighipour}, N. (2012a).
\newblock {Optical and Dynamical Characterization of Comet-like Main-belt
  Asteroid (596) Scheila}.
\newblock {\em \apj}, 744:9.

\bibitem[{Hsieh} et~al., 2012b]{HH+12}
{Hsieh}, H.~H., {Yang}, B., {Haghighipour}, N., {Kaluna}, H.~M., {Fitzsimmons},
  A., {Denneau}, L., {Novakovi{\'c}}, B., {Jedicke}, R., {Wainscoat}, R.~J.,
  {Armstrong}, J.~D., {Duddy}, S.~R., {Lowry}, S.~C., {Trujillo}, C.~A.,
  {Micheli}, M., {Keane}, J.~V., {Urban}, L., {Riesen}, T., {Meech}, K.~J.,
  {Abe}, S., {Cheng}, Y.-C., {Chen}, W.-P., {Granvik}, M., {Grav}, T., {Ip},
  W.-H., {Kinoshita}, D., {Kleyna}, J., {Lacerda}, P., {Lister}, T., {Milani},
  A., {Tholen}, D.~J., {Vere{\v s}}, P., {Lisse}, C.~M., {Kelley}, M.~S.,
  {Fern{\'a}ndez}, Y.~R., {Bhatt}, B.~C., {Sahu}, D.~K., {Kaiser}, N.,
  {Chambers}, K.~C., {Hodapp}, K.~W., {Magnier}, E.~A., {Price}, P.~A., and
  {Tonry}, J.~L. (2012b).
\newblock {Discovery of Main-belt Comet P/2006 VW$_{139}$ by Pan-STARRS1}.
\newblock {\em \apjl}, 748:L15.

\bibitem[{Ishiguro} et~al., 2011a]{Ish11a}
{Ishiguro}, M., {Hanayama}, H., {Hasegawa}, S., {Sarugaku}, Y., {Watanabe},
  J.-i., {Fujiwara}, H., {Terada}, H., {Hsieh}, H.~H., {Vaubaillon}, J.~J.,
  {Kawai}, N., {Yanagisawa}, K., {Kuroda}, D., {Miyaji}, T., {Fukushima}, H.,
  {Ohta}, K., {Hamanowa}, H., {Kim}, J., {Pyo}, J., and {Nakamura}, A.~M.
  (2011a).
\newblock {Observational Evidence for an Impact on the Main-belt Asteroid (596)
  Scheila}.
\newblock {\em \apjl}, 740:L11.

\bibitem[{Ishiguro} et~al., 2011b]{Ish11b}
{Ishiguro}, M., {Hanayama}, H., {Hasegawa}, S., {Sarugaku}, Y., {Watanabe},
  J.-i., {Fujiwara}, H., {Terada}, H., {Hsieh}, H.~H., {Vaubaillon}, J.~J.,
  {Kawai}, N., {Yanagisawa}, K., {Kuroda}, D., {Miyaji}, T., {Fukushima}, H.,
  {Ohta}, K., {Hamanowa}, H., {Kim}, J., {Pyo}, J., and {Nakamura}, A.~M.
  (2011b).
\newblock {Interpretation of (596) Scheila's Triple Dust Tails}.
\newblock {\em \apjl}, 741:L24.

\bibitem[{Ivezi{\'c}} et~al., 2001]{Ive01}
{Ivezi{\'c}}, {\v Z}., {Tabachnik}, S., {Rafikov}, R., {Lupton}, R.~H.,
  {Quinn}, T., {Hammergren}, M., {Eyer}, L., {Chu}, J., {Armstrong}, J.~C.,
  {Fan}, X., {Finlator}, K., {Geballe}, T.~R., {Gunn}, J.~E., {Hennessy},
  G.~S., {Knapp}, G.~R., {Leggett}, S.~K., {Munn}, J.~A., {Pier}, J.~R.,
  {Rockosi}, C.~M., {Schneider}, D.~P., {Strauss}, M.~A., {Yanny}, B.,
  {Brinkmann}, J., {Csabai}, I., {Hindsley}, R.~B., {Kent}, S., {Lamb}, D.~Q.,
  {Margon}, B., {McKay}, T.~A., {Smith}, J.~A., {Waddel}, P., {York}, D.~G.,
  and {the SDSS Collaboration} (2001).
\newblock {Solar System Objects Observed in the Sloan Digital Sky Survey
  Commissioning Data}.
\newblock {\em \aj}, 122:2749--2784.

\bibitem[{Jacobson} and {Scheeres}, 2011]{JaS11}
{Jacobson}, S.~A. and {Scheeres}, D.~J. (2011).
\newblock {Dynamics of rotationally fissioned asteroids: Source of observed
  small asteroid systems}.
\newblock {\em \icarus}, 214:161--178.

\bibitem[{Jehin} et~al., 2011]{Jeh11}
{Jehin}, E., {Gillon}, M., {Queloz}, D., {Magain}, P., {Manfroid}, J.,
  {Chantry}, V., {Lendl}, M., {Hutsem{\'e}kers}, D., and {Udry}, S. (2011).
\newblock {TRAPPIST: TRAnsiting Planets and PlanetesImals Small Telescope}.
\newblock {\em The Messenger}, 145:2--6.

\bibitem[{Jewitt}, 2012]{Jew12}
{Jewitt}, D. (2012).
\newblock {The Active Asteroids}.
\newblock {\em \aj}, 143:66.

\bibitem[{Jewitt} et~al., 2013]{Jew13}
{Jewitt}, D., {Agarwal}, J., {Weaver}, H., {Mutchler}, M., and {Larson}, S.
  (2013).
\newblock {The Extraordinary Multi-tailed Main-belt Comet P/2013 P5}.
\newblock {\em \apjl}, 778:L21.

\bibitem[{Jewitt} et~al., 2011]{Jew11}
{Jewitt}, D., {Stuart}, J.~S., and {Li}, J. (2011).
\newblock {Pre-discovery Observations of Disrupting Asteroid P/2010 A2}.
\newblock {\em \aj}, 142:28.

\bibitem[{Jewitt} et~al., 2010]{Jew10}
{Jewitt}, D., {Weaver}, H., {Agarwal}, J., {Mutchler}, M., and {Drahus}, M.
  (2010).
\newblock {A recent disruption of the main-belt asteroid P/2010A2}.
\newblock {\em \nat}, 467:817--819.

\bibitem[{Jewitt} et~al., 2009]{Jew09}
{Jewitt}, D., {Yang}, B., and {Haghighipour}, N. (2009).
\newblock {Main-Belt Comet P/2008 R1 (Garradd)}.
\newblock {\em \aj}, 137:4313--4321.

\bibitem[{Kim} et~al., 2012]{Kim12}
{Kim}, J., {Ishiguro}, M., {Hanayama}, H., {Hasegawa}, S., {Usui}, F.,
  {Yanagisawa}, K., {Sarugaku}, Y., {Watanabe}, J.-i., and {Yoshida}, M.
  (2012).
\newblock {Multiband Optical Observation of the P/2010 A2 Dust Tail}.
\newblock {\em \apjl}, 746:L11.

\bibitem[{Kleyna} et~al., 2013]{Kle13}
{Kleyna}, J., {Hainaut}, O.~R., and {Meech}, K.~J. (2013).
\newblock {P/2010 A2 LINEAR. II. Dynamical dust modelling}.
\newblock {\em \aap}, 549:A13.

\bibitem[{Kresak}, 1980]{Kre80}
{Kresak}, L. (1980).
\newblock {Dynamics, interrelations and evolution of the systems of asteroids
  and comets}.
\newblock {\em Moon and Planets}, 22:83--98.

\bibitem[Landolt, 1992]{landolt92}
Landolt, A. (1992).
\newblock Ubvri photometric standard stars in the magnitude range 11.5 $<$ v
  $<$ 16.0 around the celestial equator.
\newblock {\em Astrophys. J.}, 104:340--371.

\bibitem[{Licandro} et~al., 2011]{Lic11}
{Licandro}, J., {Campins}, H., {Tozzi}, G.~P., {de Le{\'o}n}, J.,
  {Pinilla-Alonso}, N., {Boehnhardt}, H., and {Hainaut}, O.~R. (2011).
\newblock {Testing the comet nature of main belt comets. The spectra of
  133P/Elst-Pizarro and 176P/LINEAR}.
\newblock {\em \aap}, 532:A65.

\bibitem[{Lowry} et~al., 2007]{Low07}
{Lowry}, S.~C., {Fitzsimmons}, A., {Pravec}, P., {Vokrouhlick{\'y}}, D.,
  {Boehnhardt}, H., {Taylor}, P.~A., {Margot}, J.-L., {Gal{\'a}d}, A., {Irwin},
  M., {Irwin}, J., and {Kusnir{\'a}k}, P. (2007).
\newblock {Direct Detection of the Asteroidal YORP Effect}.
\newblock {\em Science}, 316:272--.

\bibitem[{Marzari} et~al., 2011]{Mar11}
{Marzari}, F., {Rossi}, A., and {Scheeres}, D.~J. (2011).
\newblock {Combined effect of YORP and collisions on the rotation rate of small
  Main Belt asteroids}.
\newblock {\em \icarus}, 214:622--631.

\bibitem[{Meech} and {Svoren}, 2004]{MeS04}
{Meech}, K.~J. and {Svoren}, J. (2004).
\newblock {\em {Using cometary activity to trace the physical and chemical
  evolution of cometary nuclei}}, pages 317--335.
\newblock {Festou, M.~C., Keller, H.~U., \& Weaver, H.~A.}

\bibitem[{Moreno} et~al., 2011]{Mor11}
{Moreno}, F., {Lara}, L.~M., {Licandro}, J., {Ortiz}, J.~L., {de Le{\'o}n}, J.,
  {Al{\'{\i}}-Lagoa}, V., {Ag{\'{\i}}s-Gonz{\'a}lez}, B., and {Molina}, A.
  (2011).
\newblock {The Dust Environment of Main-Belt Comet P/2010 R2 (La Sagra)}.
\newblock {\em \apjl}, 738:L16.

\bibitem[{Moreno} et~al., 2012]{Mor12}
{Moreno}, F., {Licandro}, J., and {Cabrera-Lavers}, A. (2012).
\newblock {A Short-duration Event as the Cause of Dust Ejection from Main-Belt
  Comet P/2012 F5 (Gibbs)}.
\newblock {\em \apjl}, 761:L12.

\bibitem[{Pecaut} and {Mamajek}, 2013]{Pec13}
{Pecaut}, M.~J. and {Mamajek}, E.~E. (2013).
\newblock {Intrinsic Colors, Temperatures, and Bolometric Corrections of
  Pre-main-sequence Stars}.
\newblock {\em \apjs}, 208:9.

\bibitem[{Pravec} et~al., 2002]{Pra02}
{Pravec}, P., {Harris}, A.~W., and {Michalowski}, T. (2002).
\newblock {Asteroid Rotations}.
\newblock {\em Asteroids III}, pages 113--122.

\bibitem[{Prialnik} and {Rosenberg}, 2009]{PrR09}
{Prialnik}, D. and {Rosenberg}, E.~D. (2009).
\newblock {Can ice survive in main-belt comets? Long-term evolution models of
  comet 133P/Elst-Pizarro}.
\newblock {\em \mnras}, 399:L79--L83.

\bibitem[{Richardson} et~al., 2005]{Ric05}
{Richardson}, D.~C., {Elankumaran}, P., and {Sanderson}, R.~E. (2005).
\newblock {Numerical experiments with rubble piles: equilibrium shapes and
  spins}.
\newblock {\em \icarus}, 173:349--361.

\bibitem[{Schorghofer}, 2008]{Sch08}
{Schorghofer}, N. (2008).
\newblock {The Lifetime of Ice on Main Belt Asteroids}.
\newblock {\em \apj}, 682:697--705.

\bibitem[{Snodgrass}, 2013]{Sno13}
{Snodgrass}, C. (2013).
\newblock {Are Main Belt Comets driven by water ice sublimation?}
\newblock {\em EPSC Abstracts}, 8:EPSC2013--927.

\bibitem[{Snodgrass} et~al., 2008]{Sno08}
{Snodgrass}, C., {Saviane}, I., {Monaco}, L., and {Sinclaire}, P. (2008).
\newblock {EFOSC2 Episode IV: A New Hope}.
\newblock {\em The Messenger}, 132:18--19.

\bibitem[{Snodgrass} et~al., 2010]{Sno10}
{Snodgrass}, C., {Tubiana}, C., {Vincent}, J.-B., {Sierks}, H., {Hviid}, S.,
  {Moissi}, R., {Boehnhardt}, H., {Barbieri}, C., {Koschny}, D., {Lamy}, P.,
  {Rickman}, H., {Rodrigo}, R., {Carry}, B., {Lowry}, S.~C., {Laird}, R.~J.~M.,
  {Weissman}, P.~R., {Fitzsimmons}, A., {Marchi}, S., and {OSIRIS Team} (2010).
\newblock {A collision in 2009 as the origin of the debris trail of asteroid
  P/2010A2}.
\newblock {\em \nat}, 467:814--816.

\bibitem[{Stevenson} et~al., 2012]{Ste12}
{Stevenson}, R., {Kramer}, E.~A., {Bauer}, J.~M., {Masiero}, J.~R., and
  {Mainzer}, A.~K. (2012).
\newblock {Characterization of Active Main Belt Object P/2012 F5 (Gibbs): A
  Possible Impacted Asteroid}.
\newblock {\em \apj}, 759:142.

\bibitem[{Taylor} and {Margot}, 2013]{Tay13}
{Taylor}, P.~A. and {Margot}, J.-L. (2013).
\newblock {Tidal End States of Binary Asteroid Systems with a Nonspherical
  Component}.
\newblock {\em ArXiv e-prints}.

\bibitem[{Taylor} et~al., 2007]{Tay07}
{Taylor}, P.~A., {Margot}, J.-L., {Vokrouhlick{\'y}}, D., {Scheeres}, D.~J.,
  {Pravec}, P., {Lowry}, S.~C., {Fitzsimmons}, A., {Nolan}, M.~C., {Ostro},
  S.~J., {Benner}, L.~A.~M., {Giorgini}, J.~D., and {Magri}, C. (2007).
\newblock {Spin Rate of Asteroid (54509) 2000 PH5 Increasing Due to the YORP
  Effect}.
\newblock {\em Science}, 316:274--.

\bibitem[{Tody}, 1986]{IRAF}
{Tody}, D. (1986).
\newblock {The IRAF data reduction and analysis system}.
\newblock In {\em ``SPIE Instrumentation in Astronomy VI,'' Ed. D. L.
  Crawford}, volume 627, pages 733--748.

\bibitem[{Tsinganis} et~al., 2005]{Tsi05}
{Tsinganis}, K., {Gomes}, R., {Morbidelli}, A., and {Levison}, H. (2005).
\newblock {Origin of the Orbital Architecture of the Giant Planets of the Solar
  System}.
\newblock {\em Nature}, 435:459--61.

\bibitem[{Walsh} et~al., 2011]{Wal11}
{Walsh}, K., {Morbidelli}, A., {Raymond}, S., {O’Brien}, D.~P., and
  {Mandell}, A.~M. (2011).
\newblock {A Low Mass for Mars From Jupiter’s Early Gas-Driven Migration}.
\newblock {\em Nature}, 475:206--209.

\end{thebibliography}

\end{document}